\documentclass[acmlarge]{acmart}
\AtBeginDocument{%
  }

\setcopyright{acmlicensed}
\copyrightyear{2025}
\acmYear{2025}
\acmDOI{XXXXXXX.XXXXXXX}

\acmJournal{JOCCH}
\acmVolume{37}
\acmNumber{4}
\acmArticle{111}
\acmMonth{8}

\begin{document}

\title{Named Entity Recognition of Historical Text via Large Language Models}

\author{Shibingfeng Zhang}
\affiliation{%
 \institution{University of Bologna}
 \city{Bologna}
 \country{Italy}}

\author{Giovanni Colavizza}
\affiliation{%
 \institution{University of Bologna}
 \city{Bologna}
 \country{Italy}}
 \affiliation{%
 \institution{University of Copenhagen}
 \city{Copenhagen}
 \country{Denmark}}

\renewcommand{\shortauthors}{}

\begin{abstract}
Large language models (LLMs) have demonstrated remarkable versatility across a wide range of natural language processing tasks and domains. One such task is Named Entity Recognition (NER), which involves identifying and classifying proper names in text, such as people, organizations, locations, dates, and other specific entities. NER plays a crucial role in extracting information from unstructured textual data, enabling downstream applications such as information retrieval from unstructured text.

Traditionally, NER is addressed using supervised machine learning approaches, which require large amounts of annotated training data. However, historical texts present a unique challenge, as the annotated datasets are often scarce or nonexistent, due to the high cost and expertise required for manual labeling.  In addition, the variability and noise inherent in historical language, such as inconsistent spelling and archaic vocabulary, further complicate the development of reliable NER systems for these sources.

In this study, we explore the feasibility of applying LLMs to NER in historical documents using zero-shot and few-shot prompting strategies, which require little to no task-specific training data. Our experiments, conducted on the HIPE-2022 (Identifying Historical People, Places and other Entities)~\cite{ehrmann2022extended} dataset, show that LLMs can achieve reasonably strong performance on NER tasks in this setting. While their performance falls short of fully supervised models trained on domain-specific annotations, the results are nevertheless promising. These findings suggest that LLMs offer a viable and efficient alternative for information extraction in low-resource or historically significant corpora, where traditional supervised methods are infeasible.
\end{abstract}




\maketitle

\section{Introduction} 
Named Entity Recognition (NER) is a foundational task in natural language processing (NLP) that involves identifying entities such as people, organizations, and locations in the text under examination. Typically, given a textual input, the system is required to detect spans of entities that belong to a predefined set of categories and identify the entities' type. It serves as a crucial component in various NLP applications, including information extraction, recommendation systems, and question answering. Traditionally, building effective NER systems requires domain-specific datasets annotated with named entities. These systems, whether based on classical machine learning or deep learning, are trained using features or embeddings derived from the annotated data.

In the context of historical texts, NER can significantly aid both scholarly research and public access to archival materials. For example, a historian or researcher may be interested in tracing the appearances of a particular public figure or institution across a collection of newspapers spanning several decades. In such cases, effective entity recognition is essential for enabling structured exploration of large corpora consists of unstructured texts. While modern NER systems have achieved high accuracy on contemporary texts of various domains, applying these systems to historical documents presents unique challenges due to various reasons. Historical texts often contain archaic language, inconsistent spelling, limited punctuation, and various types of noise introduced by digitization processes such as optical character recognition. Furthermore, annotated resources for historical NER task are typically scarce due to the high cost and and expertise required for annotation, making it more difficult to apply supervised methods that require large amounts of annotated training data of high quality~\cite{ehrmann2023named}.

Recent years have witnessed the rise of Large language models (LLMs) and their versatility. LLMs refers to language models that are characterized by large-scale parameters that are trained on large quantity of text in an unsupervised manner. These language models have demonstrated many emerging capabilities not possessed by traditional task-specific models. Typically, to perform a NER task, a machine learning model must be trained and evaluated on a dataset with NER annotations from the same domain. This requirement makes it impossible to carry out the NER task without annotated data. Nevertheless, LLMs have shown the ability to overcome the limitation of annotation resources and to produce high-quality predictions with few or even zero annotated examples~\cite{keraghel2024survey}. Several studies have demonstrated that LLMs can perform NER effectively even with minimal or no annotated data at all, and various strategies have been proposed to further enhance their performance in low-resource scenarios~\cite{xie2024self,zhouuniversalner}.

Motivated by the potential of LLMs and the current limitations of NER in historical texts, this study aims to investigate the feasibility of applying LLMs to NER in historical documents across multiple languages. The primary objectives are:

\begin{itemize}
\item to evaluate the effectiveness of LLMs for performing NER on historical texts in various languages and to analyze their performance;
\item to explore strategies for improving NER performance in low-resource settings, with a particular focus on evaluating different example selection methods for few-shot learning.
\end{itemize}

The main contributions of this study are:

\begin{itemize}
\item An empirical evaluation of the feasibility of performing named entity recognition on historical texts using LLMs across different languages;
\item A comparative analysis of zero-shot and few-shot strategies for applying LLMs to historical NER tasks;
\item The  evaluation of several example selection methods to enhance few-shot NER performance in historical contexts.
\end{itemize}

The code used in this paper is publicly available~\footnote{\url{https://github.com/Zhangshibf/Named-Entity-Recognition-of-Historical-Text-via-Large-Language-Model}}.

This paper is organized into seven sections. Section~\ref{related} provides an overview of the NER task and its current state in the historical text domain. Section~\ref{dataset} describes the dataset used in this study and the evaluation metrics. Section~\ref{method} describes the methods adopted for NER, including various prompt-based settings. Section~\ref{results} presents and analyzes the experimental results. Section~\ref{discussion} discusses the implications of the results, compares the performance of different prompting strategies, highlights limitations, and outlines directions for future research. Finally, Section~\ref{conclusions} summarizes the findings and discusses potential future work.

\section{Related Works}
\label{related}
Traditionally, named entity recognition (NER) is approached using supervised methods, where the parameters of a model are learned through training and evaluation on annotated datasets. With the emergence of large language models (LLMs) in recent years and their demonstrated ability to transfer learned knowledge across a wide range of tasks, many studies have begun to explore the feasibility of applying LLMs to NER in low-resource settings. NER in historical texts is one such scenario.

This section provides a comprehensive overview of historical text NER and the application of LLMs to NER. Part~\ref{related3} discusses NER approaches in the context of historical texts. Part~\ref{related1} reviews recent advances in applying LLMs to NER, with particular attention to cases involving minimal or no training data.

\subsection{NER in Historical Texts}
\label{related3}
There is no strict definition of the concept ``Historical Texts''. Typically, it could refer to any textual documents that are created before the advent of widespread digital production and preservation, including manuscripts, newspapers, and archival materials produced in earlier historical periods. Such resources are often digitized through scanning and OCR, resulting in digital transcriptions of the original texts.

NER in the domain of historical texts holds significant value, as it enables the systematic extraction of information from unstructured sources and thereby facilitates more rigorous analysis and interpretation within historical and humanities research. Nevertheless, applying NER to historical texts presents numerous challenges. The digitized versions of such texts are often noisy, due to layout recognition errors and OCR errors. In many cases, the historical texts under examination span extensive time periods, leading to substantial variation in writing style, language, and naming conventions. In addition, the annotation of historical texts requires a high level of domain expertise, rendering high-quality annotated data for training NER systems scarce~\cite{ehrmann2023named}.

There are several datasets of historical texts with NER annotations, covering different text genres such as newspaper~\cite{ruokolainen2018recherche,ehrmann2020extended}, literary text~\cite{erdmann-etal-2016-challenges}, novels~\cite{bamman-etal-2019-annotated}, etc. A variety of approaches have been proposed to address the NER task in the domain of historical texts. Early studies employed rule-based methods and traditional machine learning techniques with manually engineered features, while recent research has shifted toward supervised deep learning approaches that leverage text embeddings~\cite{ehrmann2023named}. For example, Schweter et al. \cite{schweter2022hmbert} pre-trained a multilingual BERT model on historical texts from diverse sources, including the European Library and the British Library. This model was then fine-tuned and evaluated on historical text annotated with NER information. Boros et al.~\cite{boros2022knowledge} used Sentence-BERT~\cite{reimers-gurevych-2019-sentence} to generate embeddings from external knowledge bases, including a Wikipedia dump and Wikidata. For each historical document to be processed, relevant knowledge was retrieved and encoded using Sentence-BERT. The document itself was embedded using a BERT model fine-tuned on the target dataset. The knowledge and document embeddings were then concatenated and passed through a CRF layer for final entity prediction.

Apart from the supervised deep learning methods, some recent studies have also explored the use of LLMs to address the NER task. These studies are presented and discussed in Section~\ref{related1}

\subsection{Applying LLMs on NER}
\label{related1}
Large Language Models (LLMs) refer to attention-based neural architectures characterized by their large number of parameters. These models are pre-trained on large-scale text corpora in an unsupervised or self-supervised manner. Prominent examples include the GPT series~\cite{brown2020language}, LLaMA~\cite{dubey2024llama}, and DeepSeek~\cite{liu2024deepseek}. Typically, LLMs are generative models. For the application on downstream tasks such as text classification and question answering, LLMs have demonstrated in-context learning abilities that are not possessed by smaller language models that require fine-tuning on task-specific data~\cite{naveed2023comprehensive}. In-context learning refers to the ability of LLMs to condition their outputs on examples provided within the input prompt, effectively performing new tasks without explicit parameter updates. This phenomenon challenges traditional distinctions between training and inference, as models appear to adapt to novel tasks at inference time without additional gradient-based optimization.

Different approaches have been proposed to apply LLMs to the task of NER, with very frequently a focus on methods that require little or no training. These include prompt design for zero-shot and few-shot in-context learning, fine-tuning LLMs on specific annotated datasets, and model distillation where LLM-generated NER data is used to train smaller student models. For example, Wang et al.~\cite{wang2023gptnernamedentityrecognition} proposed GPT-NER, a training-free method that reformulates NER as a text generation task to better align with the capabilities of LLMs. Xie et al.~\cite{xie2024self} propose a training-free self-improving framework for zero-shot NER. Their method uses LLMs to annotate an unlabeled corpus by verifying the self-consistency of LLM, filters reliable annotations, and then retrieves these as demonstrations for in-context learning. Zhou et al.~\cite{zhouuniversalner} proposed UniversalNER, a distillation framework that uses instruction tuning to distill ChatGPT into smaller models optimized for open-domain NER. There have been several studies that seek to apply LLM for the NER of historical text. Gonzalez et al.~\cite{gonzalez2023yes} investigated conducting NER on datasets consists of historical French newspaper in a zero-shot manner using GPT-3.5 model~\footnote{\url{https://openai.com/blog/chatgpt}}. For each dataset, they defined a simple prompt that included a list of candidate entity tags along with the sentence to be annotated. This approach has demonstrated to be effective despite its simplicity. In contrast, Tudor et al.~\cite{tudor2025prompting} experimented with zero-shot NER using a series of smaller models under 13 billion parameters, finding that models of this scale struggle significantly with the task and are generally not suitable for zero-shot NER on historical text.

Inspired by previous works, the present study explores the application of large language models to the task of named entity recognition in historical texts, a domain characterized by limited annotated data and significant linguistic variability. Building on these works, this study investigates the use of few-shot prompting with LLMs, which enables models to generalize from a small number of task-specific examples provided at inference time rather than relying on extensive supervised training or fine-tuning~\cite{min2022rethinking}. By leveraging the generalization capabilities of LLMs, the study aims to identify effective strategies for performing NER in historical corpora with minimal reliance on manual annotation. Further details on the methodologies employed will be presented in Section~\ref{method}.



\section{Dataset and Evaluation}
\label{dataset}

This section describes the dataset in the experiments and the evaluation methods.

HIPE-2022 (Identifying Historical People, Places and other Entities)~\cite{ehrmann2022extended} dataset introduced a collection of datasets for NER in historical documents, covering the period from the 18th to the 20th century. The collection consists of six datasets that covers five languages, including German, French, English, Swedish, and Finnish. The texts primarily include newspapers and historical commentaries, all of which are manually annotated with entity information for categories such as person, location, and work. Some datasets also include more fine-grained entity types and nested entity annotations. The datasets are pre-split into training, development, and test sets, with a few exceptions that include only development and test splits.

\begin{table}[ht]
\centering
\caption{Setences with coarse NER annotations from HIPE-2022}
\resizebox{\textwidth}{!}{\begin{tabular}{|l|p{7cm}|p{7cm}|}
\hline
\textbf{Dataset} & \textbf{Text} & \textbf{NER Annotation} \\
\hline
Sonar (de) & Neueste Mittheilungen. Verantwortlicher Herausgeber: Dr. H. Klee. V. Jahrgang. Berlin, Dienstag, den 22. Juni 1886. No. 67. & [(H. Klee, PERSON), (Berlin, LOCATION)] \\
\hline
AJMC (en) & In editing the Fragments, I have availed myself of Mr. R. Ellis’ acute remarks on them in the Cambridge Journal of Philology, Vol. IV, and that I am largely indebted, as every editor must now be, to the edition of the Tragic Fragments by A. Nauck, Leipzig, 1856. & [(R. Ellis, PERSON), (Cambridge Journal of Philology, WORK), (Vol. IV, SCOPE), (A. Nauck, PERSON), (Leipzig, LOCATION), (1856, DATE)] \\
\hline
\end{tabular}}

\label{table:hipe}
\end{table}

The dataset defines three tasks: (1) \textbf{coarse NER}, which uses a general set of entity categories; (2) \textbf{fine-grained NER}, which includes more specific entity types and supports nested annotations, meaning that one entity can be contained within another entity; and (3) \textbf{entity linking}. This study focuses on the coarse NER task, as it provides a foundational basis for evaluating LLM capabilities across diverse languages and historical genres, while keeping the task complexity manageable for initial benchmarking. Table~\ref{table:hipe} shows some example from HIPE-2022, with coarse NER annotations.

The HIPE-2022 dataset includes an official evaluation tool called HIPE-scorer\footnote{\url{https://github.com/hipe-eval/HIPE-scorer}}. This tool evaluates NER predictions under two different settings, namely fuzzy and strict. In the fuzzy setting, a prediction is considered correct if it overlaps with the ground truth and shares the same NER tag. In contrast, the strict setting requires a prediction to have exactly the same boundaries and tag as the ground truth to be counted as correct. In this study, results under both settings are reported to provide a more balanced assessment of the system’s performance.

\section{Methodologies}
\label{method}
This section details the methodologies adopted for performing NER on historical documents. The methodologies involved in this study cover the zero-shot method ( Section~\ref{zeroshot}), which serves as the baseline, and various few-shot methods examples retrieved using different strategies (Section~\ref{fewshot}). The LLM used in this study is DeepSeek-V3-0324~\cite{liu2024deepseek}. It was selected due to its strong performance across a variety of tasks and languages, as well as its relatively low API usage cost compared to other LLMs.

Few-shot learning refers to the ability of a LLM to generalize a task from only a small number of labeled examples. In the context of NER, this typically involves providing an LLM with a handful of annotated sentences as demonstrations in the prompt, enabling the model to adapt its predictions to the specific labeling schema without requiring full-scale supervised training~\cite{moscato2023few}. Prior work has shown that the choice of examples strongly influences performance. Examples that are semantically or structurally similar to the target input tend to yield better results than randomly chosen ones. For example, Xie et al.~\cite{xie2024self} proposed a training-free self-improving framework for zero-shot NER that leverages an unlabeled corpus to generate pseudo-labeled data via self-consistency. They then filter these predictions through reliability measures, forming a self-annotated dataset. From this dataset, examples are retrieved and used as in-context demonstrations for inference. Their experiments on multiple benchmarks demonstrate that carefully selected examples, particularly those that are both reliable and similar to the target text, significantly outperform random demonstrations. This finding motivates our retrieval-based few-shot methods, which aim to select contextually relevant examples from historical texts.

\begin{figure}[!ht]
\begin{center}
\includegraphics[width=0.6\columnwidth]{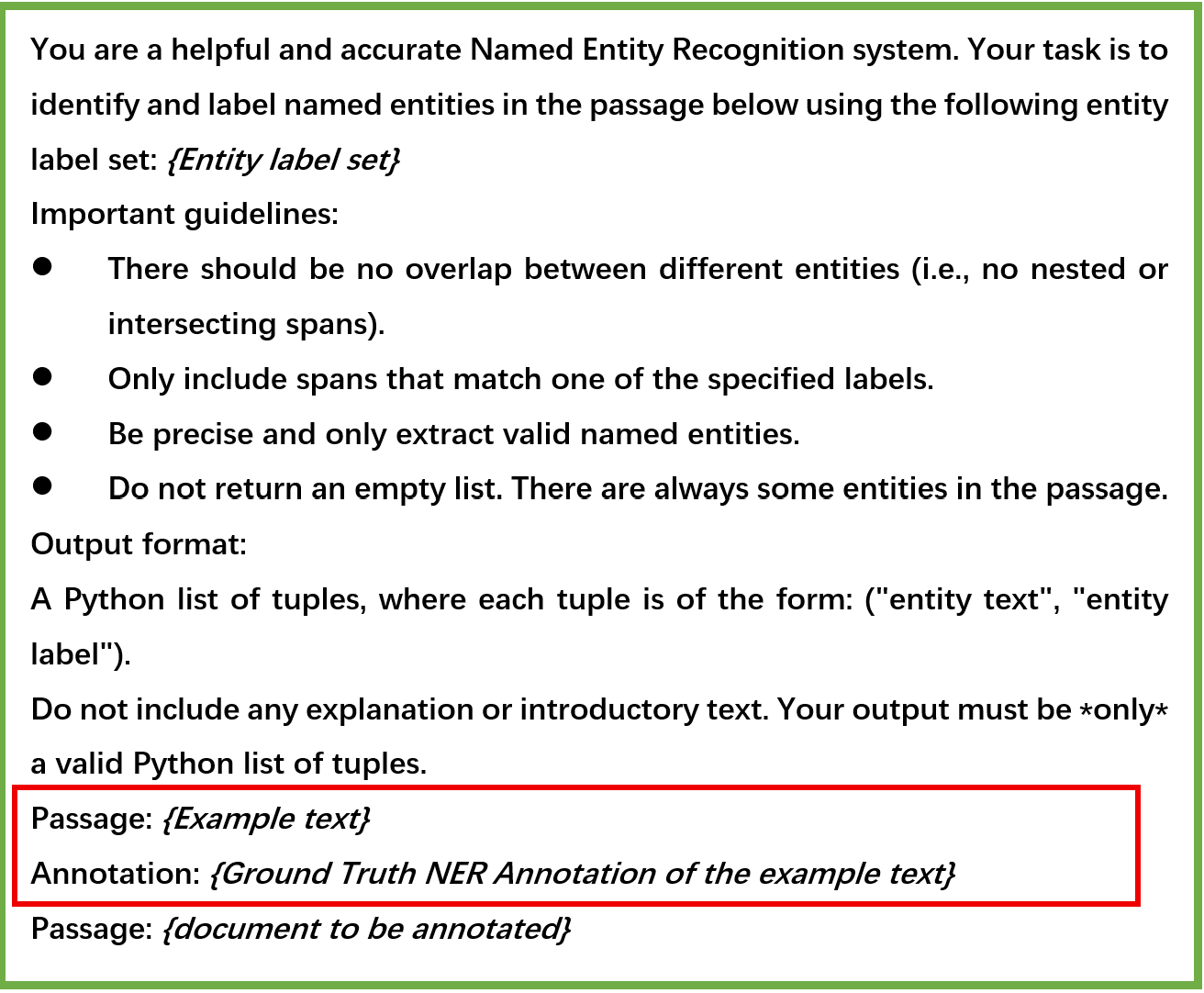}
\caption{Template used for prompt generation. Notice that content in red rectangle is applicable only to few-shot prompt.}
\label{figureprompt}
\end{center}
\end{figure}

The NER experimental procedure adopted in this study consists of the following steps:

\begin{enumerate}
\item \textbf{Example Retrieval.} Given a text to be annotated, the first step is to retrieve similar texts from the training and development sets to serve as in-context examples for the LLM. Various retrieval methods can be applied. The zero-shot baseline does not include this step. Further details of the retrival methods are provided in Section~\ref{zeroshot} and Section~\ref{fewshot}.

\item \textbf{Prompt Generation.} Using the retrieved examples and the target text, a prompt is constructed according to a predefined template and submitted to the LLM via API. Figure~\ref{figureprompt} shows the template of prompt. This process is described further in detail in Section~\ref{promptdesign}.

\item \textbf{Response Processing.} The LLM's response is converted into the IOB annotation format for evaluation and saved. This process is described with more details in Section~\ref{post}.

\end{enumerate}

The temperature of LLM is set to 0 in all experiments.  In order to account for the indeterministic nature of LLM, both experiments in zero-shot and few-shot settings are repeated three times and the mean and standard deviation of the results are reported. To further explore the self-consistency of the LLM and its application on NER task~\cite{xie2024self}, majority voting is also conducted using the response gathered from the three runs of experiments. This procedure is presented in Section~\ref{voting}.
\subsection{Prompt Design}
\label{promptdesign}
The prompt is designed following zero-shot settings or few-shot settings. Apart from the inclusion of examples retrieved from the dataset in the few-shot setting, the remainder of the prompt remains identical across both configurations.
\subsubsection{Zero-shot Settings}
\label{zeroshot}
The zero-shot method, adopted as the baseline in this study, is the simplest approach evaluated. For each document to be annotated, the LLM is provided with the text, along with instructions specifying the set of entity tags to be used and the desired output format. The prompt adopted in zero-shot setting can be found in Appendix~\ref{zero}.
\subsubsection{Few-shot Settings}
Few-shot methods are employed to further enhance the performance of the LLM on the task. The prompt template used in this setting is provided in Appendix~\ref{few}. Each example is one article from the dataset.

To retrieve examples for the few-shot setting, two strategies are applied, namely \textit{Lexical Overlap} and \textit{Embedding Similarity}. The former is designed to capture surface-level lexical similarity, while the latter is designed to capture deeper semantic similarity. For each document to be annotated, examples are always selected from a combination of the training and development sets. A random retrieval strategy is also included for comparison against other methods that select examples based on similarity between the document to be annotated and those in the training and development sets. All few-shot methods are evaluated using 1, 3, and 5 example settings.

\label{fewshot}
\paragraph{Example Selection Based on Lexical Overlap} 
\label{selection_overlap}

The lexical overlap method identifies candidate examples by measuring the token-level similarity between the target document and candidate documents, weighted by their Term Frequency-Inverse Document Frequency (TF-IDF) scores. This approach conducts surface-level matches of token across the training set and development set of each dataset. The procedure is performed following these steps:

\begin{enumerate}
    \item \textbf{Pre-processing} The text in the dataset is already in the format of tokens, therefore tokenization is not needed in this step. Stop words and punctuations are removed using stop word list from NLTK~\cite{bird-loper-2004-nltk}.

    \item \textbf{Token Filtering}
The TF-IDF score for each token is computed. To eliminate unimportant tokens, a filtering step is applied based on these scores. Specifically, for each document, tokens falling within the bottom 10\% of TF-IDF scores across the entire dataset are removed.

\item \textbf{Overlap Score Calculation}
The similarity between a target document and a candidate document, which is from the combination of the training and development sets, is calculated based on overlapping tokens, their TF-IDF scores, and their relative frequencies. This calculation is represented by the following equation:

\begin{equation}
    \text{Overlap Score}(d_t, d_c) = 
    \sum_{t \in \mathcal{T}(d_t) \cap \mathcal{T}(d_c)} 
    \left( \text{TF-IDF}_t(t) + \text{TF-IDF}_c(t) \right) 
    \times \left( \frac{\text{tf}_t(t)}{|\mathcal{T}(d_t)|} + \frac{\text{tf}_c(t)}{|\mathcal{T}(d_c)|} \right)
\end{equation}

Where $d_t$ represents the target document, $d_c$ represents the candidate document, $\mathcal{T}(d_t)$ represents the filtered tokens count of target document, $\mathcal{T}(d_c)$ represents the filtered tokens count of candidate document. The first term combines TF-IDF weights from both documents. The second term normalizes by token frequency in each document. Higher scores indicate better better matches at the token level.
\item \textbf{Ranking} Based on the overlap score, candidate documents are ranked by their similarity to the target document. The top-$k$ most similar documents are selected accordingly.
\end{enumerate}

\paragraph{Example Selection Based on Embedding Similarity}
\label{selection_embedding}
This method uses sentence embeddings to measure semantic similarity between documents via cosine similarity~\cite{reimers-gurevych-2019-sentence}. The embeddings are generated using the \textit{distiluse-base-multilingual-cased-v2} model, selected for its efficiency, multilingual support, and lightweight architecture.

Given a target document, the top-$k$ most similar candidate documents are identified by computing the cosine similarity scores between the target and each candidate.


\subsubsection{Post-processing of LLM Response}
\label{post}
The prompt instructs the LLM to return extracted entities as a Python list of tuples, where each tuple consists of two strings: the entity text as it appears in the input, and its corresponding label. To facilitate evaluation, this output is converted into the IOB format used by the HIPE-2022 dataset by matching entity substrings against the original text and mapping them to the corresponding dataset file. Tuples containing text not found in the input are discarded.

\subsection{Majority Voting}
\label{voting}
Similar to the self-consistency decoding strategy proposed by Wang et al.\cite{wang2023selfconsistency}, which improves reasoning tasks by sampling multiple diverse outputs and aggregating them into a more accurate final answer, the ensemble method in this work leverages repeated runs to reduce variance and filter out spurious predictions. This also aligns with Xie et al.\cite{xie2024self}, who demonstrate that leveraging multiple pseudo-labeled outputs and retaining only reliable ones enhances NER performance. These studies support the use of majority voting as a lightweight but effective strategy for improving the robustness of NER predictions in historical texts.

As mentioned before, in order to implement this strategy and to account for the non-deterministic nature of LLMs, each experiment is repeated three times under the same setup. A majority voting scheme is then applied to the three resulting sets of predictions. For each token, the final tag is assigned based on the majority vote across the three runs. In cases of a tie, no tag is assigned to the token.

\section{Results}
\label{results}
This section presents the experimental results. Evaluation is performed using the HIPE-scorer\footnote{\url{https://github.com/hipe-eval/HIPE-scorer}}, the official evaluation tool for the HIPE dataset, to ensure the reliability and comparability of results with previous studies.  The HIPE-scorer provides two evaluation methods: strict and fuzzy. Under the strict setting, a prediction is considered correct only if it exactly matches both the span and the label of the ground truth. Under the fuzzy setting, a prediction is considered correct if it overlaps with the ground truth and has the correct label. All experiments were repeated three times, and the reported results include the mean and estimated confidence intervals of the evaluation metrics under both the strict and fuzzy evaluation settings.

The section is organized into two parts. Part~\ref{sub1} compares and analyzes the results obtained using different prompting methods. Part~\ref{sub2} compares the best-performing results from this study with those reported in previous work.

\subsection{Evaluation of Prompting Methods}
\label{sub1}

\begin{table}[t]
  \caption{Experimental results comparing different prompting strategies. Reported values are the mean over three runs, with subscripts indicating confidence intervals estimated using the Student's t-distribution. The evaluation metric is micro Strict F1 score ($\pm$ confidence interval) for each method on each dataset. Best results per row are highlighted in bold. 
\textit{baseline} refers to the zero-shot prompt design described in Section~\ref{zeroshot}. 
\textit{r1}, \textit{r3}, and \textit{r5} represent few-shot prompting with 1, 3, and 5 randomly retrieved examples, respectively. 
\textit{embedding1}, \textit{embedding3}, and \textit{embedding5} use examples selected based on embedding similarity between target and candidate documents. 
\textit{overlap1}, \textit{overlap3}, and \textit{overlap5} use examples selected based on token-level overlap.
 }
  \label{tab:micro_strict}
  \centering
   \small
\resizebox{\textwidth}{!}{  \begin{tabular}{lcccccccccc}
    \toprule
    Dataset & baseline & r1 & r3 & r5 & embedding1 & embedding3 & embedding5 & overlap1 & overlap3 & overlap5 \\
    \midrule

ajmc (de) & 0.241$_{\pm 0.021}$ & 0.671$_{\pm 0.091}$ & 0.676$_{\pm 0.074}$ & 0.654$_{\pm 0.017}$ & 0.681$_{\pm 0.027}$ & 0.681$_{\pm 0.034}$ & 0.644$_{\pm 0.010}$ & 0.714$_{\pm 0.007}$ & \textbf{0.724$_{\pm 0.057}$} & 0.699$_{\pm 0.006}$ \\
ajmc (en) & 0.292$_{\pm 0.016}$ & 0.605$_{\pm 0.075}$ & 0.615$_{\pm 0.061}$ & 0.613$_{\pm 0.044}$ & 0.602$_{\pm 0.031}$ & 0.628$_{\pm 0.020}$ & 0.602$_{\pm 0.052}$ & \textbf{0.641$_{\pm 0.010}$} & 0.631$_{\pm 0.024}$ & 0.607$_{\pm 0.041}$ \\
ajmc (fr) & 0.403$_{\pm 0.011}$ & 0.665$_{\pm 0.021}$ & 0.678$_{\pm 0.005}$ & 0.688$_{\pm 0.021}$ & 0.679$_{\pm 0.007}$ & 0.699$_{\pm 0.023}$ & 0.675$_{\pm 0.009}$ & 0.706$_{\pm 0.008}$ & \textbf{0.726$_{\pm 0.026}$} & 0.715$_{\pm 0.033}$ \\
hipe2020 (de) & 0.459$_{\pm 0.009}$ & 0.513$_{\pm 0.021}$ & 0.481$_{\pm 0.012}$ & 0.458$_{\pm 0.005}$ & \textbf{0.525$_{\pm 0.003}$} & 0.490$_{\pm 0.017}$ & 0.484$_{\pm 0.007}$ & 0.525$_{\pm 0.013}$ & 0.499$_{\pm 0.016}$ & 0.477$_{\pm 0.015}$ \\
hipe2020 (en) & 0.516$_{\pm 0.009}$ & \textbf{0.558$_{\pm 0.028}$} & 0.544$_{\pm 0.039}$ & 0.550$_{\pm 0.029}$ & 0.527$_{\pm 0.011}$ & 0.550$_{\pm 0.005}$ & 0.546$_{\pm 0.024}$ & 0.555$_{\pm 0.014}$ & 0.547$_{\pm 0.012}$ & 0.553$_{\pm 0.008}$ \\
hipe2020 (fr) & 0.475$_{\pm 0.009}$ & 0.532$_{\pm 0.026}$ & 0.509$_{\pm 0.030}$ & 0.510$_{\pm 0.011}$ & 0.564$_{\pm 0.016}$ & 0.521$_{\pm 0.013}$ & 0.511$_{\pm 0.017}$ & \textbf{0.566$_{\pm 0.013}$} & 0.533$_{\pm 0.013}$ & 0.515$_{\pm 0.016}$ \\
letemps (fr) & 0.473$_{\pm 0.003}$ & 0.493$_{\pm 0.016}$ & 0.477$_{\pm 0.028}$ & 0.469$_{\pm 0.009}$ & \textbf{0.536$_{\pm 0.008}$} & 0.510$_{\pm 0.007}$ & 0.492$_{\pm 0.004}$ & 0.479$_{\pm 0.006}$ & 0.477$_{\pm 0.001}$ & 0.473$_{\pm 0.007}$ \\
newseye (de) & 0.367$_{\pm 0.005}$ & 0.386$_{\pm 0.027}$ & 0.377$_{\pm 0.014}$ & 0.376$_{\pm 0.013}$ & 0.390$_{\pm 0.009}$ & 0.384$_{\pm 0.009}$ & \textbf{0.393$_{\pm 0.012}$} & 0.383$_{\pm 0.008}$ & 0.366$_{\pm 0.006}$ & 0.377$_{\pm 0.013}$ \\
newseye (fi) & 0.387$_{\pm 0.019}$ & 0.438$_{\pm 0.032}$ & 0.433$_{\pm 0.033}$ & 0.424$_{\pm 0.042}$ & 0.432$_{\pm 0.030}$ & 0.416$_{\pm 0.010}$ & 0.401$_{\pm 0.030}$ & \textbf{0.440$_{\pm 0.015}$} & 0.423$_{\pm 0.034}$ & 0.417$_{\pm 0.031}$ \\
newseye (fr) & 0.465$_{\pm 0.011}$ & 0.512$_{\pm 0.024}$ & 0.498$_{\pm 0.007}$ & 0.490$_{\pm 0.005}$ & \textbf{0.521$_{\pm 0.015}$} & 0.500$_{\pm 0.004}$ & 0.496$_{\pm 0.005}$ & 0.473$_{\pm 0.004}$ & 0.484$_{\pm 0.005}$ & 0.485$_{\pm 0.005}$ \\
newseye (sv) & 0.443$_{\pm 0.011}$ & 0.525$_{\pm 0.047}$ & 0.495$_{\pm 0.011}$ & 0.490$_{\pm 0.015}$ & 0.527$_{\pm 0.020}$ & 0.508$_{\pm 0.006}$ & 0.480$_{\pm 0.016}$ & \textbf{0.555$_{\pm 0.026}$} & 0.508$_{\pm 0.027}$ & 0.474$_{\pm 0.020}$ \\
sonar (de) & 0.496$_{\pm 0.053}$ & \textbf{0.650$_{\pm 0.096}$} & 0.633$_{\pm 0.142}$ & 0.597$_{\pm 0.049}$ & 0.595$_{\pm 0.016}$ & 0.535$_{\pm 0.111}$ & 0.521$_{\pm 0.025}$ & 0.581$_{\pm 0.022}$ & 0.538$_{\pm 0.013}$ & 0.524$_{\pm 0.037}$ \\
topres19th (en) & 0.623$_{\pm 0.060}$ & 0.660$_{\pm 0.024}$ & 0.633$_{\pm 0.014}$ & 0.621$_{\pm 0.011}$ & \textbf{0.687$_{\pm 0.005}$} & 0.663$_{\pm 0.014}$ & 0.659$_{\pm 0.025}$ & 0.671$_{\pm 0.004}$ & 0.659$_{\pm 0.013}$ & 0.652$_{\pm 0.012}$ \\
 \midrule
Average & 0.434$_{\pm 0.018}$ & 0.554$_{\pm 0.041}$ & 0.542$_{\pm 0.036}$ & 0.534$_{\pm 0.021}$ & 0.559$_{\pm 0.015}$ & 0.545$_{\pm 0.021}$ & 0.531$_{\pm 0.018}$ & \textbf{0.561}$_{\pm 0.012}$ & 0.547$_{\pm 0.019}$ & 0.536$_{\pm 0.019}$ \\

    \bottomrule
  \end{tabular}}
\end{table}

\begin{table}[t]
  \caption{Experimental results comparing different prompting strategies. The evaluation metric is micro Fuzzy F1 score ($\pm$ confidence interval) for each method on each dataset. Reported values are the mean over three runs, with subscripts indicating confidence intervals estimated using the Student's t-distribution. Best results per dataset are highlighted in bold. 
\textit{baseline} refers to the zero-shot prompt design described in Section~\ref{zeroshot}. 
\textit{r1}, \textit{r3}, and \textit{r5} represent few-shot prompting with 1, 3, and 5 randomly retrieved examples, respectively. 
\textit{embedding1}, \textit{embedding3}, and \textit{embedding5} use examples selected based on embedding similarity between target and candidate documents. 
\textit{overlap1}, \textit{overlap3}, and \textit{overlap5} use examples selected based on token-level overlap.
}
  \label{tab:micro_fuzzy}
  \centering
   \small
\resizebox{\textwidth}{!}{  \begin{tabular}{lcccccccccc}
    \toprule
    Dataset & baseline & r1 & r3 & r5 & embedding1 & embedding3 & embedding5 & overlap1 & overlap3 & overlap5 \\
    \midrule

ajmc (de) & 0.392$_{\pm 0.043}$ & 0.743$_{\pm 0.072}$ & 0.759$_{\pm 0.055}$ & 0.740$_{\pm 0.046}$ & 0.759$_{\pm 0.019}$ & 0.751$_{\pm 0.032}$ & 0.715$_{\pm 0.026}$ & 0.782$_{\pm 0.007}$ & \textbf{0.783$_{\pm 0.024}$} & 0.767$_{\pm 0.011}$ \\
ajmc (en) & 0.459$_{\pm 0.006}$ & 0.710$_{\pm 0.075}$ & 0.708$_{\pm 0.027}$ & 0.714$_{\pm 0.046}$ & 0.700$_{\pm 0.021}$ & 0.730$_{\pm 0.019}$ & 0.708$_{\pm 0.025}$ & \textbf{0.744$_{\pm 0.008}$} & 0.729$_{\pm 0.027}$ & 0.716$_{\pm 0.034}$ \\
ajmc (fr) & 0.489$_{\pm 0.018}$ & 0.783$_{\pm 0.014}$ & 0.795$_{\pm 0.011}$ & 0.804$_{\pm 0.011}$ & 0.802$_{\pm 0.003}$ & 0.814$_{\pm 0.009}$ & 0.802$_{\pm 0.017}$ & 0.808$_{\pm 0.004}$ & \textbf{0.828$_{\pm 0.022}$} & 0.822$_{\pm 0.021}$ \\
hipe2020 (de) & 0.584$_{\pm 0.007}$ & \textbf{0.643$_{\pm 0.032}$} & 0.606$_{\pm 0.012}$ & 0.591$_{\pm 0.007}$ & 0.642$_{\pm 0.002}$ & 0.612$_{\pm 0.023}$ & 0.616$_{\pm 0.006}$ & 0.640$_{\pm 0.015}$ & 0.618$_{\pm 0.005}$ & 0.606$_{\pm 0.015}$ \\
hipe2020 (en) & 0.649$_{\pm 0.022}$ & 0.685$_{\pm 0.013}$ & 0.678$_{\pm 0.030}$ & 0.688$_{\pm 0.014}$ & 0.680$_{\pm 0.005}$ & 0.677$_{\pm 0.005}$ & 0.674$_{\pm 0.017}$ & \textbf{0.695$_{\pm 0.012}$} & 0.682$_{\pm 0.008}$ & 0.682$_{\pm 0.010}$ \\
hipe2020 (fr) & 0.633$_{\pm 0.006}$ & 0.677$_{\pm 0.019}$ & 0.654$_{\pm 0.010}$ & 0.651$_{\pm 0.011}$ & \textbf{0.703$_{\pm 0.018}$} & 0.663$_{\pm 0.015}$ & 0.649$_{\pm 0.014}$ & 0.697$_{\pm 0.011}$ & 0.669$_{\pm 0.013}$ & 0.659$_{\pm 0.012}$ \\
letemps (fr) & 0.560$_{\pm 0.004}$ & 0.581$_{\pm 0.024}$ & 0.556$_{\pm 0.034}$ & 0.543$_{\pm 0.001}$ & \textbf{0.610$_{\pm 0.005}$} & 0.576$_{\pm 0.006}$ & 0.560$_{\pm 0.008}$ & 0.562$_{\pm 0.014}$ & 0.550$_{\pm 0.004}$ & 0.547$_{\pm 0.001}$ \\
newseye (de) & 0.513$_{\pm 0.015}$ & 0.518$_{\pm 0.023}$ & 0.516$_{\pm 0.017}$ & 0.512$_{\pm 0.012}$ & 0.519$_{\pm 0.010}$ & 0.519$_{\pm 0.009}$ & \textbf{0.527$_{\pm 0.002}$} & 0.514$_{\pm 0.015}$ & 0.492$_{\pm 0.006}$ & 0.508$_{\pm 0.012}$ \\
newseye (fi) & 0.581$_{\pm 0.016}$ & \textbf{0.628$_{\pm 0.020}$} & 0.619$_{\pm 0.035}$ & 0.612$_{\pm 0.026}$ & 0.617$_{\pm 0.020}$ & 0.600$_{\pm 0.006}$ & 0.594$_{\pm 0.036}$ & 0.597$_{\pm 0.026}$ & 0.601$_{\pm 0.030}$ & 0.595$_{\pm 0.022}$ \\
newseye (fr) & 0.634$_{\pm 0.004}$ & 0.663$_{\pm 0.015}$ & 0.646$_{\pm 0.016}$ & 0.641$_{\pm 0.003}$ & \textbf{0.665$_{\pm 0.016}$} & 0.645$_{\pm 0.005}$ & 0.642$_{\pm 0.004}$ & 0.642$_{\pm 0.004}$ & 0.647$_{\pm 0.018}$ & 0.648$_{\pm 0.000}$ \\
newseye (sv) & 0.664$_{\pm 0.013}$ & 0.718$_{\pm 0.017}$ & 0.697$_{\pm 0.009}$ & 0.705$_{\pm 0.017}$ & 0.705$_{\pm 0.012}$ & 0.706$_{\pm 0.012}$ & 0.695$_{\pm 0.003}$ & \textbf{0.724$_{\pm 0.025}$} & 0.707$_{\pm 0.014}$ & 0.681$_{\pm 0.012}$ \\
sonar (de) & 0.616$_{\pm 0.026}$ & \textbf{0.761$_{\pm 0.086}$} & 0.735$_{\pm 0.114}$ & 0.720$_{\pm 0.056}$ & 0.755$_{\pm 0.012}$ & 0.668$_{\pm 0.056}$ & 0.657$_{\pm 0.018}$ & 0.728$_{\pm 0.036}$ & 0.662$_{\pm 0.042}$ & 0.652$_{\pm 0.044}$ \\
topres19th (en) & 0.684$_{\pm 0.047}$ & 0.720$_{\pm 0.022}$ & 0.685$_{\pm 0.019}$ & 0.674$_{\pm 0.017}$ & \textbf{0.743$_{\pm 0.009}$} & 0.715$_{\pm 0.011}$ & 0.711$_{\pm 0.031}$ & 0.733$_{\pm 0.004}$ & 0.711$_{\pm 0.026}$ & 0.706$_{\pm 0.010}$ \\

 \midrule
Average & 0.574$_{\pm 0.017}$ & 0.679$_{\pm 0.033}$ & 0.666$_{\pm 0.030}$ & 0.661$_{\pm 0.021}$ & \textbf{0.685}$_{\pm 0.012}$ & 0.667$_{\pm 0.016}$ & 0.658$_{\pm 0.016}$ & 0.682$_{\pm 0.014}$ & 0.668$_{\pm 0.018}$ & 0.661$_{\pm 0.016}$ \\

    \bottomrule
  \end{tabular}}
\end{table}

\begin{table}[h]
\centering
\small
\caption{Experiment results of the performance of majority voting using predictions from three runs. The evaluation metrics are Strict Micro F1 and Fuzzy Micro F1. \textit{Best Performance} refers to the highest scores from Tables~\ref{tab:micro_fuzzy} and~\ref{tab:micro_strict}. \textit{Best Voted} refers to the best results achieved through majority voting across the three runs. \textit{Vote Gain} refers to the difference between Best Voted and Best Performance.}
\label{tab:voted}
\begin{tabular}{l r r r r r r}

\toprule
Dataset & Best Performance & Best Voted & Vote Gain & Best Performance & Best Voted & Vote gain \\
 & \multicolumn{3}{c}{Strict Micro F1} & \multicolumn{3}{c}{Fuzzy Micro F1} \\
\cmidrule(lr){2-4} \cmidrule(lr){5-7}
ajmc (de)& 0.724 & 0.729 & +0.006 & 0.783 & 0.794 & +0.011 \\
ajmc (en)& 0.641 & 0.645 & +0.004 & 0.744 & 0.748 & +0.004 \\
ajmc (fr)& 0.726 & 0.731 & +0.004 & 0.828 & 0.832 & +0.004 \\
hipe2020 (de)& 0.525 & 0.524 & -0.001 & 0.643 & 0.646 & +0.002 \\
hipe2020 (en)& 0.558 & 0.560 & +0.002 & 0.695 & 0.704 & +0.009 \\
hipe2020 (fr)& 0.566 & 0.571 & +0.005 & 0.703 & 0.708 & +0.005 \\
letemps (fr)& 0.536 & 0.538 & +0.002 & 0.610 & 0.613 & +0.002 \\
newseye (de)& 0.393 & 0.393 & +0.000 & 0.527 & 0.530 & +0.003 \\
newseye (fi)& 0.440 & 0.449 & +0.009 & 0.628 & 0.624 & -0.004 \\
newseye (fr)& 0.521 & 0.520 & -0.002 & 0.665 & 0.670 & +0.005 \\
newseye (sv)& 0.555 & 0.560 & +0.004 & 0.724 & 0.726 & +0.002 \\
sonar (de)& 0.650 & 0.640 & -0.010 & 0.761 & 0.762 & +0.002 \\
topres19th (en)& 0.687 & 0.688 & +0.001 & 0.743 & 0.743 & +0.000 \\
 \midrule
 Average& 0.579&0.581&0.002&0.696&0.700&0.003\\
\bottomrule
\end{tabular}

\end{table}
This Section compares the results achieved by different prompting methods. The HIPE-2022 datasets are relatively limited in size, with test sets often comprising only around twenty documents. Such small evaluation sets can lead to high variance and unreliable comparisons. To address this issue, and considering that the proposed method does not require training, all experiments comparing prompting strategies report the results on the combined training and development sets. This setup allows for more robust and stable evaluation.

Table~\ref{tab:micro_strict} presents the experimental results on HIPE-2022 using strict F1 scores and and corresponding confidence intervals. Table~\ref{tab:micro_fuzzy} reports the results using fuzzy F1 scores and corresponding confidence intervals.

As shown in both tables, a consistent pattern that can be observed is that prompt methods with in-context examples achieve better performance across all datasets, demonstrating the effectiveness of providing examples. Even adding a single random example to the prompt yields better results than the baseline across all datasets, demonstrating the benefits of in-context learning. Another observation is that prompt methods using only one example outperform their counterparts using three or five examples across almost all datasets, this is likely due to the increased prompt length introduced by multiple examples, which may have lead to exceeding the model’s optimal context window~\cite{liu-etal-2024-lost}. Further Wilcoxon tests with Bonferroni correction demonstrate that all few-shot prompting methods significantly outperform the zero-shot baseline for both strict and fuzzy F1 scores. Among the few-shot approaches, \textit{r1} significantly outperforms both \textit{r3} and \textit{r5}, while \textit{overlap1} significantly outperforms \textit{overlap3} and \textit{overlap5}, confirming that single-example prompts are more effective than multi-example variants. Similarly, \textit{embedding1} significantly outperforms \textit{embedding5}, with no significant difference found between \textit{embedding1} and \textit{embedding3} for strict F1, though \textit{embedding1} shows significant superiority for fuzzy F1. Notably, no significant differences were detected between the three single-example methods (\textit{r1}, \textit{embedding1}, and \textit{overlap1}) on either fuzzy F1 score or strict F1 score, indicating that the choice of example selection strategy has less impact on performance than simply providing a single in-context example versus multiple examples or no examples at all.

Further error analysis is conducted to better understand the benefits of different prompting strategies. In the HIPE-2022 scorer's evaluation framework, the errors are categorized into 5 types:
\begin{enumerate}
\label{entity types}
    \item Spurious entity error: an entity is predicted, but it does not exist in the ground truth.
    \item Missed entity error: an entity exists in the ground truth but is not predicted.
    \item Entity type substitution error: the predicted span matches the ground-truth span, but the predicted entity type is incorrect.
    \item Entity span substitution error: the predicted entity label is correct, but the predicted span only partially overlaps with the true span. Note that this case is considered an error only in strict evaluation settings, not in fuzzy evaluation settings.
    \item Entity type and span substitution error: the predicted entity label is incorrect, and the predicted span only partially overlaps with the true span.
\end{enumerate}

\begin{figure}[!ht]
\begin{center}
\includegraphics[width=0.6\columnwidth]{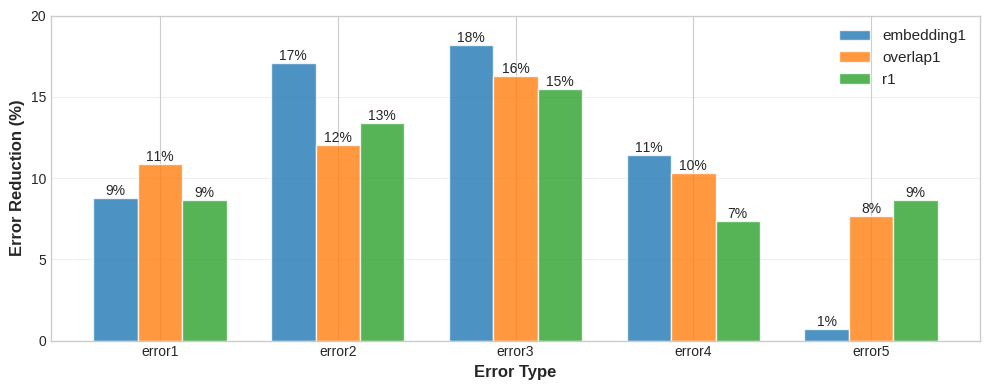}
\caption{Error reduction rate of the embedding1, overlap1, and r1 systems in comparison to the baseline. For an explanation of the system names, see the caption of Table~\ref{tab:micro_strict}. For the definitions of error types, see List~\ref{entity types}.}
\label{fig.1}
\end{center}
\end{figure}

We calculated the error reduction rates of the three best-performing systems compared to the baseline across all datasets. As shown in Figure~\ref{fig.1}, the methods exhibit similar patterns: all three systems are most effective at reducing type 2 errors (failing to predict an existing entity) and type 3 errors (assigning the wrong label to a detected entity). This suggests that exposure to an example helps the models better align entity recognition with the correct categories and improves their sensitivity to entity presence.

\begin{figure}[!ht]
\begin{center}
\includegraphics[width=0.6\columnwidth]{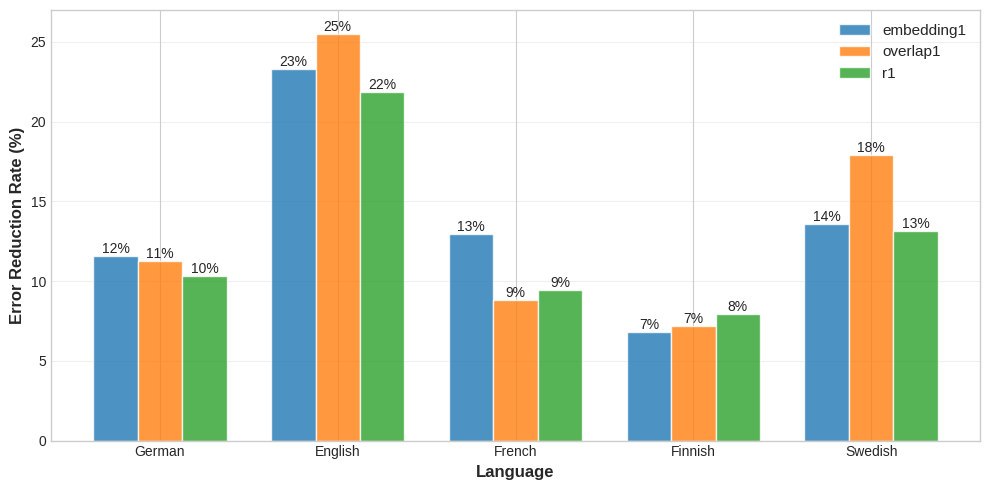}
\caption{Language-wise error reduction rate of the embedding1, overlap1, and r1 systems in comparison to the baseline. For an explanation of the system names, see the caption of Table~\ref{tab:micro_strict}. For the definitions of error types, see List~\ref{entity types}.}
\label{fig.2}
\end{center}
\end{figure}

To examine the impact across different languages, we computed the error reduction rates for the same systems on different languages. Figure~\ref{fig.2} shows that providing an example is particularly effective for English datasets, while its impact is smaller for less resource-rich languages.

To improve performance, majority voting was applied over the predictions of three runs. The results are reported in Table~\ref{tab:voted}. As shown, majority voting generally leads to improved performance, although the gains are often modest. Further Wilcoxon test indicates that the gains from majority voting are statistically insignificant under the strict evaluation setting but significant under the fuzzy setting.
\subsection{Comparison with Existing Results}
\label{sub2}

\begin{table}[t]
  \caption{Comparison between state-of-the-art results and the best performance achieved in the present study.  \textit{SOTA} refers to the current state-of-the-art, while \textit{Best Performance} refers to the highest performance achieved in this study, which are the mean over three runs, with subscripts indicating confidence intervals estimated using the Student's t-distribution.}

  \label{tab:paper}
  \centering
   \small
    \begin{tabular}{lcccccc}
    \toprule
   Dataset & SOTA & Best & $\Delta$ Strict & SOTA & Best & $\Delta$ Fuzzy \\
            & \multicolumn{3}{c}{Strict Micro F1} & \multicolumn{3}{c}{Fuzzy Micro F1} \\
    \cmidrule(lr){2-4} \cmidrule(lr){5-7}
ajmc (de) & 0.934~\cite{boros2022knowledge}& 0.728$_{\pm 0.016}$ &-0.206&0.952~\cite{boros2022knowledge} & 0.769$_{\pm 0.024}$ &-0.183\\
ajmc (en) & 0.877~\cite{gonzalez2023injecting}& 0.657$_{\pm 0.066}$ &-0.22& 0.933~\cite{gonzalez2023injecting}& 0.754$_{\pm 0.088}$ &-0.179\\
ajmc (fr) & 0.844~\cite{gonzalez2023injecting}& 0.717$_{\pm 0.077}$ &-0.127&0.897~\cite{gonzalez2023injecting} & 0.821$_{\pm 0.037}$ &-0.076\\
hipe2020 (de) &0.794~\cite{boros2022knowledge} & 0.579$_{\pm 0.027}$ &-0.215&0.876~\cite{boros2022knowledge} & 0.690$_{\pm 0.018}$ &-0.186\\
hipe2020 (en) & 0.630~\cite{gonzalez2023injecting}& 0.561$_{\pm 0.024}$ &-0.069&0.777~\cite{gonzalez2023injecting} & 0.699$_{\pm 0.017}$ &-0.078\\
hipe2020 (fr) &0.808~\cite{boros2022knowledge} & 0.595$_{\pm 0.011}$ &-0.213& 0.907~\cite{boros2022knowledge}& 0.727$_{\pm 0.014}$ &-0.18\\
letemps (fr) & 0.661~\cite{ehrmann2022extended}& 0.533$_{\pm 0.076}$ &-0.128& 0.711~\cite{ehrmann2022extended}& 0.603$_{\pm 0.069}$ &-0.108\\
newseye (de) &0.477~\cite{ehrmann2022extended} & 0.326$_{\pm 0.008}$ &-0.151&0.570~\cite{ehrmann2022extended} & 0.420$_{\pm 0.017}$ &-0.15\\
newseye (fi) & 0.644~\cite{ehrmann2022extended}& 0.516$_{\pm 0.080}$ &-0.128& 0.760~\cite{ehrmann2022extended}& 0.674$_{\pm 0.083}$ &-0.086\\
newseye (fr) & 0.656~\cite{ryser2022exploring}& 0.487$_{\pm 0.042}$ &-0.169&0.786~\cite{ryser2022exploring} & 0.619$_{\pm 0.037}$ &-0.167\\
newseye (sv) & 0.651~\cite{ehrmann2022extended}& 0.566$_{\pm 0.052}$ &-0.085& 0.747~\cite{ehrmann2022extended}& 0.691$_{\pm 0.042}$ &-0.056\\sonar (de) & 0.529~\cite{ryser2022exploring}& 0.580$_{\pm 0.040}$ &0.051& 0.695~\cite{ryser2022exploring}& 0.717$_{\pm 0.072}$ &0.022\\
topres19th (en) &0.787~\cite{ryser2022exploring} & 0.709$_{\pm 0.003}$ &-0.078& 0.838~\cite{ryser2022exploring} & 0.752$_{\pm 0.005}$ &-0.086\\

    \bottomrule
  \end{tabular}
\end{table}

Table~\ref{tab:paper} compares the state-of-the-art (SOTA) results on the HIPE-2022 datasets with those achieved in this paper. The external papers referenced in the table employ supervised deep learning approaches, involving training or fine-tuning language models on the training set of the corresponding dataset or other specialized datasets. In contrast, the methods explored in this study leverage LLM in zero-shot or few-shot settings, without any fine-tuning or additional training on the task-specific data. 

As shown in Table~\ref{tab:paper}, there is a clear performance gap between supervised SOTA methods and LLM-based approaches. This is expected, as the SOTA models benefit from direct exposure to the training data, allowing them to learn dataset-specific patterns and annotation conventions. The only dataset where the proposed method outperforms the reported SOTA is Sonar. While the performance of LLM-based methods does not yet match supervised approaches, these results highlight their promise as a cost-effective and language-agnostic baseline for historical NER. These approaches are particularly valuable for low-resource or multilingual settings where annotated data is scarce or unavailable.

Overall, the experiments demonstrate that few-shot prompting with a single example provides the most consistent improvements over the zero-shot baseline, regardless of the example selection strategy. While majority voting further improve results, its impact is modest and mainly observed under the fuzzy evaluation setting. Compared to fully supervised state-of-the-art approaches, LLM-based prompting methods still underperform, with the only exception of the Sonar dataset, but they offer a flexible, cost-efficient, and multilingual alternative for historical NER tasks, especially in low-resource scenarios where annotated training data is not available.

\section{Discussion}
\label{discussion}
The findings of this study provide several insights into the behavior of LLM-based prompting approaches for historical NER. Experiments were conducted using zero-shot and few-shot strategies. For few-shot prompting, different example selection methodologies were tested, including random selection, selection based on lexical similarity, and selection based on embedding similarity. Each few-shot strategy was evaluated with 1, 3, and 5 examples provided in the prompt.

The results show that few-shot prompting with a single example consistently outperforms the zero-shot baseline across all datasets. This highlights the remarkable effectiveness of minimal in-context learning, demostrating that even a single example is sufficient to improve the model's performance. Counterintuitively, providing more examples tends to decrease performance. This is likely due to longer prompts exceeding the model’s optimal context window or diluting the clarity of the task description, a phenomenon also reported in recent work on in-context learning~\cite{liu-etal-2024-lost}. The analysis also shows that the method of selecting examples (random selection, embedding-based selection, or lexical overlap-based selection) has less influence on results than the presence of an example itself. While lexical overlap and embedding-based retrieval occasionally yield slight improvements, the differences are not statistically significant compared to random selection. This indicates that LLMs are able to generalize from minimal demonstrations regardless of how they are chosen. This finding raises questions about how to design more targeted and efficient retrieval strategies that could exploit corpus-specific features.

Majority voting over multiple runs improves performance, particularly under fuzzy evaluation. However, the observed gains are generally modest, and under strict evaluation they do not reach statistical significance. This suggests that ensembling at the prediction level is not sufficient to overcome the inherent variability of prompting methods on small evaluation sets.

In general, LLM-based prompting on HIPE-2022 dataset remains behind supervised SOTA systems on nearly all datasets. This performance gap is unsurprising, as SOTA systems benefit from training directly on annotated corpora, allowing them to adapt to dataset-specific annotation schemes. The only exception is the Sonar dataset, where LLM-based prompting slightly outperforms prior work. The general trend confirms that prompting cannot yet replace supervised fine-tuning when high accuracy is essential.

The practical implications of these results are nonetheless significant. Prompting approaches offer a cost-effective, language-agnostic, and training-free alternative for historical NER. In scenarios where annotated data is scarce, expensive to produce, or unavailable, prompting provides a viable baseline. This aligns with the increasing emphasis in digital humanities and historical linguistics on methods that can scale across multilingual and under-resourced datasets without extensive manual annotation. At the same time, the study highlights important limitations. The HIPE-2022 test sets are relatively small, sometimes consisting of only a few dozen documents, which introduces high variance and limits the robustness of statistical comparisons. Furthermore, the experiments were restricted to a single LLM, and results may differ across models with varying architectures, context lengths, and training data.

Future work should therefore explore prompt optimization, including compression techniques and the development of automatic methods for constructing concise yet informative prompts. More sophisticated retrieval strategies, potentially leveraging semantic and historical metadata, could also improve the quality of in-context examples. Lastly, evaluating across broader historical corpora and additional languages would help assess the generalizability of these findings and strengthen the role of LLMs in supporting multilingual historical research.
\section{Conclusions}
\label{conclusions}
This study explored the feasibility of performing named entity recognition on historical texts using large language models in a training-free approach. Prior work has shown that LLMs can perform NER effectively with little or no annotated data and that few-shot learning is efficient in such scenarios~\cite{xie2024self, wang2023selfconsistency, zhouuniversalner}. Based on these findings, a series of experiments was conducted using the HIPE-2022 dataset, a multilingual historical corpus with NER annotations. The experiments evaluated LLM performance under zero-shot and few-shot settings. The zero-shot settings serve as the baseline. The few-shot settings cover different example retrieval strategies, including random retrieval, lexical-based retrieval, and embedding-based retrieval, with varying numbers of in-context examples.

The results indicate that few-shot prompting consistently improves performance over zero-shot baselines, even with a single in-context example. Increasing the number of examples beyond one often led to diminished performance, likely due to longer prompts exceeding the model’s optimal context window. The choice of example selection strategy had a minimal impact on outcomes, suggesting that LLMs are capable of generalizing from minimal demonstrations. While majority voting over multiple runs slightly improves fuzzy evaluation scores, gains remain modest and shows no statistically significant benefit. Overall, LLM-based prompting performs below supervised state-of-the-art systems on most datasets, with the exception of the Sonar dataset where it slightly surpasses prior work.

Despite these limitations, the study demonstrates the practical potential of LLM prompting for historical NER. Training-free prompting provides a cost-effective, language-agnostic alternative in scenarios where annotated data is scarce, costly, or unavailable, supporting research in multilingual and under-resourced historical corpora. In conclusion, while prompting cannot yet replace supervised fine-tuning for high-accuracy historical NER, it offers a viable baseline and a flexible tool for researchers in digital humanities. The results encourage further exploration of methods that combine minimal supervision with LLM capabilities to advance historical text analysis.
\begin{acks}
To be added
\end{acks}
  
\bibliographystyle{ACM-Reference-Format}
\bibliography{sample-base}
 \appendix

\section{Prompt used in zero-shot setting} \label{zero}

\fbox{\begin{minipage}{\textwidth}
Your task is to identify and label named entities in the passage below using the following entity label set: \textit{Entity label set}

Important guidelines:
\begin{itemize}
    \item There should be no overlap between different entities (i.e., no nested or intersecting spans).
    \item Only include spans that match one of the specified labels.
    \item Be precise and only extract valid named entities.
    \item Do not return an empty list. There are always some entities in the passage.
\end{itemize}

Output format:

A Python list of tuples, where each tuple is of the form: ("entity text", "entity label").

Do not include any explanation or introductory text. Your output must be *only* a valid Python list of tuples.

Passage: \textit{document to be annotated}

\end{minipage}}
\section{Prompt used in few-shot setting} \label{few}

\fbox{\begin{minipage}{\textwidth}
Your task is to identify and label named entities in the passage below using the following entity label set: \textit{Entity label set}

Important guidelines:
\begin{itemize}
    \item There should be no overlap between different entities (i.e., no nested or intersecting spans).
    \item Only include spans that match one of the specified labels.
    \item Be precise and only extract valid named entities.
    \item Do not return an empty list. There are always some entities in the passage.
\end{itemize}

Output format:

A Python list of tuples, where each tuple is of the form: ("entity text", "entity label").

Do not include any explanation or introductory text. Your output must be *only* a valid Python list of tuples.

Passage: \textit{Example text}

Annotation: \textit{Ground Truth NER Annotation of the example text}

Passage: \textit{document to be annotated}
\end{minipage}}

\end{document}